\documentclass[journal]{IEEEtran}

\usepackage[pdftex]{graphicx}
\usepackage[cmex10]{amsmath}
\usepackage{amssymb}
\usepackage{url}
\usepackage{booktabs}
\usepackage{multirow}
\usepackage{diagbox}
\usepackage{hyperref}
\usepackage[style=ieee,backend=biber,citestyle=numeric-comp]{biblatex}
\usepackage{tikz} \usetikzlibrary{calc, shadings, 3d, perspective, fit, overlay-beamer-styles, mindmap, positioning, shapes.geometric, decorations.markings, shapes.callouts,decorations.pathmorphing, matrix, svg.path}
\usepackage{pgfplots, pgfplotstable}\pgfplotsset{compat=newest, /pgf/number format/assume math mode=true}
\usepgfplotslibrary{groupplots, fillbetween, patchplots, colormaps, colorbrewer, dateplot, statistics}
\usepackage[linesnumbered,ruled,vlined]{algorithm2e}

\newcommand{\bx}{\mathbf{x}}
\newcommand{\bz}{\mathbf{z}}
\newcommand{\bW}{\mathbf{W}}
\newcommand{\bD}{\mathbf{D}}
\newcommand{\bd}{\mathbf{d}}
\newcommand{\bL}{\boldsymbol{\Lambda}}
\newcommand{\bO}{\boldsymbol{\Omega}}

\title{Whittaker-Henderson smoother for long satellite image time series interpolation}
\author{Mathieu~Fauvel, \IEEEmembership{Senior Member, IEEE}}

\usepackage{siunitx}
\begin{document}
\maketitle
\begin{abstract}
  Whittaker smoother is a widely adopted solution to pre-process satellite image time series. Yet, two key limitations remain: the smoothing parameter must be tuned individually for each pixel, and the standard formulation assumes homoscedastic noise, imposing uniform smoothing across the temporal dimension. This paper addresses both limitations by casting the Whittaker smoother as a differentiable neural layer, in which the smoothing parameter is inferred by a neural network. The framework is further extended to handle heteroscedastic noise through a time-varying regularization, allowing the degree of smoothing to adapt locally along the time series. To enable large-scale processing, a sparse, memory-efficient, and fully differentiable implementation is proposed, exploiting the symmetric banded structure of the underlying linear system via Cholesky factorization. Benchmarks on GPU demonstrate that this implementation substantially outperforms standard dense linear solvers, both in speed and memory consumption. The approach is validated on SITS acquired over the French metropolitan territory between 2016 and 2024. Results confirm the feasibility of large-scale heteroscedastic Whittaker smoothing, though reconstruction differences with the homoscedastic baseline remain limited, suggesting that the transformer architecture used for smoothing parameter estimation may lack the temporal acuity needed to capture abrupt noise variations such as single-day cloud contamination.
\end{abstract}

\section{Introduction}
\label{sec:intro}
Since its introduction in the remote sensing community in 2011 by Atzberger and Eilers~\cite{Atzberger10072011}, Whittaker smoother~\cite{Eilers2003} has been used in many remote sensing works to smooth or interpolate long satellite image time series (SITS). Among conventional signal processing techniques such as Fourrier filtering, local polynomial regression (e.g. Savitzky-Golay filter) or parametric models (e.g. assymetric Gaussian), Whittaker smoother, and extended versions, usually shows best results for different scenario: for instance in land cover mapping~\cite{SHAO2016258} or in gap filling of remote sensing products~\cite{bg-10-4055-2013,KONG201913,ijgi12060214,Atkinson2012400,rs12203383,Atzberger01092011}. 

Across applications, the Whittaker smoother is valued because of its simplicity and efficiency. The method uses a penalized least-squares formulation, eliminating the need for complex model fitting, and achieves good scalability with increasing data volumes through a sparse representation of the fitting problem~\cite{Eilers2003,hpfilter:efficient:computation}. This approach naturally extends to multivariate time series with a limited increase in complexity. While the original formulation was proposed to regular time series, it has since been adapted to handle irregular and unaligned SITS. In such case, observations are unevenly spaced in time, due to cloud cover or shadows, and may vary in size. Furthermore, it requires tuning of a single smoothing parameter which control the trade-off between fidelity to raw observations and smoothness of the result. Thanks to its flexibility, Whittaker smoother has been incorporated into operational remote sensing toolkits, such as the \emph{Copernicus Data Space Ecosystem}\footnote{\url{https://marketplace-portal.dataspace.copernicus.eu/catalogue/app-details/8}}.

The choice of smoothing parameter critically affects performance in terms of interpolation error. A higher value increases smoothing but risks distorting the true phenological patterns, while a lower value may preserve those patterns but leave residual noise. Original strategy was to used leave-one-out cross-validation based grid search~\cite{Eilers2003}. More sophisticated methods such as \mbox{V-curve}~\cite{8076705} or spectral entropy~\cite{bernalarencibia2025optimalsmoothingparametereilerswittaker} have then been successively investigated. Recently,~\cite{biessy:hal-04124043} has proposed a Bayesian view of the Whittaker smoother which allowed to select the smoothing parameters based on the maximization of a marginal likelihood.

These methods perform for one pixel time series, but the smoothing parameter needs to be tuned for every sample. In the context of large scale processing, it might be too demanding to tune this parameter to all pixels. A conventional strategy is to use the same smoothing value for all pixels of a given location. However, for pixels with heterogeneous dynamics (e.g., with different land cover) this strategy might fail. Additionally, the original Whittaker smoother model is based on the assumption of homoscedastic noise in the time series—meaning the noise level remains constant throughout. As a result, the degree of smoothing applied is uniform across the series. This assumption is too strong for real remote sensing time series, since atmospheric conditions vary over time and unevenly affect the original signal.

In this work, we address these two limitations by introducing a regularization neural network that dynamically infers the smoothing value for each pixel. We further adapt the regularization parameter to handle heteroscedastic noise, enabling the degree of smoothing to vary across the time dimension. Our approach leverages the analytical solution of the Whittaker smoother, integrating it as a neural layer that can be optimized through standard stochastic gradient descent and autodifferentiation. After training on large-scale datasets, the estimation of smoothing values becomes amortized—requiring only a single forward pass during inference.

The next section recalls the Whittaker objective function, introduces heteroscedastic regularization and then how the neural layer is defined. Section~\ref{sec:implementation} presents a memory efficient implementation of the sparse inverse problem. Experimental results on real long SITS over France are presented in~\ref{sec:simu}. The paper concludes with Section~\ref{conclusion}. 

\section{Whittaker smoother as a neural layer}
\label{sec:method}
Throughout this paper, \(\bx\in\mathbb{R}^{T}\) is a vector of size \(T\) containing the reflectance at times \(\{t_{1}, \ldots, t_{T}\}=\mathbf{t}\) of the sample, \(\bW\in\mathbb{R}^{T\times T}\) is a diagonal matrix where each diagonal entry \(w_{ii}\) equals 1 if the sample at time \(t_i\) has been observed and is not flagged by its associated cloud mask, otherwise it is 0, \(\bD^{k+1}\in\mathbb{R}^{(T-k-1)\times T}\) is the order \(k + 1\) discrete difference operator and \(\lambda\in\mathbb{R}_{>0}\) is the smoothing parameter. The variable \(\bW\) serves to handle unaligned time series by allowing padding in temporal dimension and to interpolated to unseen times. 

For the discrete difference operator on unevenly spaced time sample, the definition proposed in~\cite[Section 6]{10.1561/2200000099} and implement in \url{https://github.com/glmgen/dspline} was used in this work. In what follows, the case of one pixel with only one spectral band is discussed for brevity, but the framework naturally extends to batched multivariate vectors.  

\subsection{Whittaker smoother for homoscedastic time series}
\label{sec:w:smoother}
Using notation above, the Whittaker smoother solves the following problem
\begin{eqnarray}
  \label{eq:loss}
  \arg\min_{\mathbf{z}, \lambda}\Big[(\mathbf{x} - \mathbf{z})^\top\mathbf{W}(\mathbf{x} - \mathbf{z}) + \lambda\|\mathbf{D}^{(k+1)}\mathbf{z}\|^2 \Big].
\end{eqnarray}
The penalization is the L2 norm of the \(k+1\) discrete derivative of the reconstructed time series: The smaller the norm value, the smoother the function. Assuming the smoothing parameter fixed, the solution can be found explicitly  by setting the gradient w.r.t. \(\bz\) to zeros:
\begin{eqnarray}
  \label{eq:sol}
  {\mathbf{z}} = \Big[\mathbf{W} + \lambda\mathbf{D}^{(k+1)^\top}\mathbf{D}^{(k+1)}\Big]^{-1}\mathbf{W}\mathbf{x}.
\end{eqnarray}
In eq.~(\ref{eq:loss}), having a scalar value for the penalization amounts to put equal penalization for each time sample of the derivative when calculating its norm. Therefore, the reconstructed time series maintains a uniform degree of smoothness throughout the time domain, which corresponds to assuming homoscedastic (constant-variance) noise in the original time series.

\subsection{Extension to heteroscedastic time series}
\begin{figure}
  \centering
  \begin{tikzpicture}
    \begin{groupplot}[group style={group size=2 by 1, horizontal sep=50pt}, grid, footnotesize, xmin=0, xmax=15, xlabel=\(t\),]
      \nextgroupplot[width=0.575\linewidth, height=0.35\linewidth, ylabel=\(x(t)\)]
      \addplot[only marks, mark size=2pt, opacity=0.125] table[x=doy, y=data_noisy, col sep =comma] {figures/values_data.csv};
      \addplot[black, dashed] table[x=doy, y=data, col sep =comma] {figures/values_data.csv};
      \addplot[very thick, blue, ] table[x=doy, y=z_homo, col sep =comma] {figures/values_data.csv};
      \addplot[very thick, red] table[x=doy, y=z_hetero, col sep =comma] {figures/values_data.csv};
      \nextgroupplot[ymode=log,width=0.425\linewidth, height=0.35\linewidth, ylabel=\(\bL_{tt}\)]
      \addplot[very thick, blue] table[x=doy, y=alpha_0, col sep=comma] {figures/values_reg.csv};
      \addplot[very thick, red] table[x=doy, y=alpha_1, col sep=comma] {figures/values_reg.csv};
    \end{groupplot}
  \end{tikzpicture}
  \caption{Synthetic exemple of a time series with heteroscedastic noise. The dashed black curve is the true function, the light gray circles are the noisy observation and the blue/red curves are the reconstruction using the Whittaker smoother. The noise level is higher between time 6 and 10 and the Whittaker smoother with constant regularization value leads to noisy interpolation (blue curve), while the smoother with a time-varying regularization value leads to a smoother reconstruction (red curve). The right plot shows the regularization value a function of the time: for the heteroscedastic model, a piecewise constant function was used.}
  \label{fig:hpfilter:homo:hetero}
\end{figure}
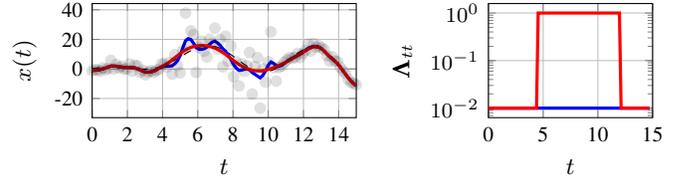

As highlighted in the introduction, this assumption is quite restrictive and frequently fails to hold in practical, real-world settings. To cope with this limitation, it is proposed to use a weighted norm for the penalization, defined as \(\|\mathbf{v}\|^2_{\bL} = \mathbf{v}^\top\bL\mathbf{v}\) with \(\mathbf{v}^\top\) the transpose of \(\mathbf{v}\), and \(\bL\) a diagonal matrix with strictly positive entries. With this setting, the weighted norm writes \(\|\mathbf{v}\|^2_{\bL} = \sum_{t=1}^T\bL_{tt}\mathbf{v}_t^2\) allowing each time sample of the derivative to have a  specific smoothing value. The solution of the Whittaker smoother problem becomes
\begin{eqnarray}
  \label{eq:sol:hetero}
  {\mathbf{z}} = \Big[\mathbf{W} + \mathbf{D}^{(k+1)^\top}\bL\mathbf{D}^{(k+1)}\Big]^{-1}\mathbf{W}\mathbf{x}.
\end{eqnarray}
Figure~\ref{fig:hpfilter:homo:hetero} shows an illustration of smoothing obtained with homoscedastic and heteroscedastic models on a synthetic time series with heteroscedastic noise.

The estimation of \(T\) values for each pixel would be infeasible in practice in terms of computation time. The proposed solution in this work is to learn a neural network to estimate the regularization values based on \(\mathbf{t}\) and \(\bx\) for all the training pixels, and to amortize its estimation for inference time. This strategy is discussed in the next section.

\subsection{Linear layer}
\label{sec:layer}
Equation~(\ref{eq:sol:hetero}) can be cast as a linear layer \(f\) in a standard neural network setting \({\bz} = f_{\bL}(\bx | \bW, \bD)\) with \(\bL=g_{\boldsymbol{\theta}}(\bx, \mathbf{t})\) and \(g\) is a neural network with parameters \(\boldsymbol{\theta}\), to be optimized. As long as the derivative of \(f\) with respect to \(\bL\) or \(\bx\) is well-defined and computable, it becomes possible to optimize \(\boldsymbol{\theta}\) using stochastic gradient descent and automatic differentiation within a deep learning framework~\cite{blondel2025elementsdifferentiableprogramming}. Fortunately, by applying standard results for matrix derivatives, the following results hold true for all \(t\in\mathbf{t}\)~\cite{IMM2012-03274}:
\begin{align}
  \frac{\partial \bz}{\partial \lambda_t} &= -\Big[\mathbf{W} + \mathbf{D}^{(k+1)^\top}\bL\mathbf{D}^{(k+1)}\Big]^{-1} \bd_t^{(k+1)}\bd_t^{(k+1)^\top}\bz   \label{eq:layer:derivative} \\
  \frac{\partial \bz}{\partial \bx} &=\bW\Big[\mathbf{W} + \mathbf{D}^{(k+1)^\top}\bL\mathbf{D}^{(k+1)}\Big]^{-1} \label{eq:layer:derivative:x}
\end{align}
with \(\bd_t^{(k+1)}\) is the \(t\)th row of \(\mathbf{D}^{(k+1)}\).

From eq.~(\ref{eq:sol:hetero}), eq.~(\ref{eq:layer:derivative}) and eq.~(\ref{eq:layer:derivative:x}), a naive implementation of forward-mode and reverse/backward-mode automatic differentiation would be computationally dominated by the need to solve linear systems: once during the forward pass and \(T\) times during the backward pass. The computational cost would scale as \(O(T^3)\) and the memory requirements  as \(O(T^2)\), for each sample or pixel, posing a significant challenge when processing large amount of pixels.

While general sparse linear algebra solvers exist, they are either non-differentiable or rely on iterative solvers, which are computationally prohibitive for large-scale applications like processing millions of pixels. Below, we demonstrate how leveraging the symmetric banded structure of the linear system enables a sparse, efficient, and differentiable implementation of the layer \(f\), fully compatible with GPU acceleration.

\section{Efficient implementation of the inverse problem}
\label{sec:implementation}
In the following, we denote \(\bO = \mathbf{W} + \mathbf{D}^{(k+1)^\top}\bL\mathbf{D}^{(k+1)}\). It is known that \(\bO\) is a symmetric positive definite band matrix with bandwidth \((k+1)\)~\cite{biessy:hal-04124043}. Following sections show how to implement a differentiable efficient solver based on this specific structure.

\subsection{Band storage and Cholesky solver}
\label{sec:banded:storage}
Band storage~\cite{anderson1999lapack} allows a compact storage of the data values reducing the amount of data to store from by considering only non zero values below the main diagonal, as illustrated in Fig.~\ref{fig:pack:storage}. Since \((k+1)\ll T\)  this approach significantly reduces memory usage from \(O(T^2)\) to \(O\left(T(k+1)\right)\).

The positive-definite property enables an efficient numerical solution of the linear system via Cholesky factorization, followed by forward and backward substitution~\cite[Chap. 4.3.5]{doi:10.1137/1.9781421407944}\footnote{A conjugate gradient solver was also investigated using the band storage and an efficient dot product. Although a good pre-conditioner was found it was less efficient in terms of processing time than the Cholesky factorisation.}. Using band storage, the Cholesky decomposition can be computed in \(O\left(T(k+1)^2\right)\) rather than \(O(T^3)\)  and the triangular system can be solved using back-substitution in \(O(T(k+1))\) instead of \(O(T^2)\)~\cite[Chap. 4.3.6]{doi:10.1137/1.9781421407944}. A pseudo code for Cholesky decomposition is given in Algorithm~\ref{alg:banded_cholesky_nobatch}. The two \emph{for} loops over \(T\) and \((k+1)\) elements exhibit the reduced complexity in \(O\left(T(k+1)^2\right)\). 

\begin{figure}
  \centering
  \begin{tikzpicture}[font=\footnotesize]
    \matrix (full) [
    matrix of math nodes,
    left delimiter=(, right delimiter=),
    row sep = -2mm, column sep=-1mm%
    ]
    {
      \omega_{11} & \omega_{21} & 0  & 0 \\
      \omega_{21} & \omega_{22} & \omega_{23} & 0 \\
      0  & \omega_{32} & \omega_{33} & \omega_{43}\\
      0 & 0 &\omega_{43} & \omega_{44} \\
    };
    \matrix (packed) [right=of full,
    matrix of math nodes,
    left delimiter=(, right delimiter=),
    row sep = -2mm, column sep=-1mm
    ]
    {
      \omega_{11} & \omega_{22} & \omega_{33} & \omega_{44} & \omega_{55} \\
      \omega_{21} & \omega_{23} & \omega_{34} & 0 & 0 \\
    };
    \begin{scope}[node distance=1ex]
      \node [above=-0.1cm of full] {Full Storage};
      \node [above=-0.1cm of packed] {Lower Band Storage};
    \end{scope}
  \end{tikzpicture}
  \caption{Illustration of the band storage for \(T=4\) and \((k+1)=1\) for \(\bO\). The left side of the figure illustrates the conventional full matrix storage, while the right side shows the band storage format, which stores only the diagonal and lower diagonal elements. As for triangular matrix, upper and lower format exists. In this work, the lower format was used.}
  \label{fig:pack:storage}
\end{figure}

\begin{algorithm}
\caption{\small In place Cholesky factorization of a symmetric banded matrix using band storage. Indices notation are given in NumPy indexing syntax.}
\label{alg:banded_cholesky_nobatch}
\footnotesize
\SetKwInOut{Input}{Input}
\Input{\(\boldsymbol{\Omega}\) in a lower band storage shown in~\ref{fig:pack:storage}}
  \For{$t = 0$ \KwTo $T$}{
    $\omega = \bO[0,t]$ \\
    $\mathbf{v} =  \bO[\mathrel{{1}{:}}, t]$\\
    $\bO[0,t] = \sqrt{\omega}$\\
    $\bO[\mathrel{{1}{:}},t] = \mathbf{v} / \sqrt{\omega}$\\
    \For{$u = 0$ \KwTo $\min\big(k+1, T-t-1\big)$}{
      $\bO[\mathrel{{:}{k}{+}{1}{-}{u}},\mathrel{{t}{+}{1}{+}{u}}]\mathrel{{-}{=}} \mathbf{v}[\mathrel{{u}{:}{u}{+}{k}{+}{1}}]\mathbf{v}[u]/ \omega $
    }   
  }
\end{algorithm}
\subsection{Implementation}
\label{sec:torch:implementation}
Band storage and Cholesky solver have been implemented in plain Pytorch~\cite{10.5555/3454287.3455008} using its facility to extend autograd operator\footnote{\url{https://docs.pytorch.org/tutorials/beginner/examples_autograd/polynomial_custom_function.html}}. It allows to benefit from high CPU/GPU computing efficiency and to optimize the parameters of \(g_{\boldsymbol{\theta}}\) using all the available stochastic gradient machinery. The code is available here \url{https://src.koda.cnrs.fr/mathieu.fauvel.4/hp-filter}.

The proposed implementation of Whittaker smoother has been compared with a standard implementation using Torch solver\footnote{\url{https://docs.pytorch.org/docs/stable/generated/torch.linalg.solve.html}}. Batch of multi-variate time series of length \(T=350\) and number of feature \(C=10\) were generated using Sentinel-2 pixels (full data description is given in section~\ref{sec:dataset}). Several batch sizes were generated and the computation were done on a \emph{GPU Tesla V100 32 Go}. For the band storage, different order for the discrete derivative were investigated since the complexity of the Cholesky decomposition and the forward/backward substitution are related to this parameter. Five runs were done and averaged results are reported in Table~\ref{tab:bench}. The layer was set in \emph{train} mode, i.e., the computational graph for the forward and backward pass were computed.

From the table, it can be seen that the full storage solver does not beyond 8192 pixels because of memory issue. Hence, even for small batch size, the proposed implementation provides substantial speed-up. Additionally, it scales effectively to larger batch sizes, as expected, due to the efficient band storage format. For large batch sizes, the order of the band matrix significantly impacts processing time, doubling the computation time when increasing the order from 2 to 4. Yet, the processing time for a batch size of 28672 for order 4 is still more than 3 times lower than the conventional solver on a batch size of 4096. This implementation enables application of both homoscedastic and heteroscedastic Whittaker smoothers to long and large-scale SITS, with results presented in the following section.

\begin{table}
  \centering
  \caption{Processing time (in second) to solve eq.~(\ref{eq:sol:hetero}) for different batch size. A \(\emptyset\) indicates that the algorithm run out of memory.}
  \label{tab:bench}
  \scriptsize
  {
    \addtolength{\tabcolsep}{-1pt} 
  \begin{tabular}{ccccccccc}
    \toprule
   \textbf{Solver} &\textbf{Order} & \multicolumn{7}{c}{\textbf{Batch size}}\\
    & & 4096 & 8192 & 12288 & 16384 & 20480 & 24576 & 28672 \\
    \midrule
  \multirow{3}{*}{Cholesky} & 2 & 0.14 & 0.14 & 0.14 & 0.13 & 0.14 & 0.13 & 0.14 \\
                             & 3 & 0.17 & 0.15 & 0.16 & 0.15 & 0.19 & 0.19 & 0.20 \\
                             & 4 & 0.18 & 0.18 & 0.20 & 0.18 & 0.21 & 0.24 & 0.27 \\
  \midrule
  Full                       &   & 0.49 & 1.07 & \(\emptyset\) & \(\emptyset\) & \(\emptyset\) & \(\emptyset\) & \(\emptyset\) \\
  \bottomrule
\end{tabular}}
\end{table}

\section{Experimental results}
\label{sec:simu}
\subsection{Dataset}
\label{sec:dataset}
Data were downloaded from the Theia Data Center\footnote{Theia Data Center}. 19 tiles over the French metropolitan territory have been selected. Surface reflectance time-series were produced using the MAJA processing chain, which corrects atmospheric, adjacency, and slope effects, and provides cloud and shadow masks~\cite{baetens_validation_2019}. These masks was used to define valid and non valid temporal acquisitions. All spectral bands at a spatial resolution of 10 and 20 m/pixel were used. Bands at 20 m/pixel were spatially upsampled to 10m/pixel using \emph{sensorio} python library (\url{https://github.com/CNES/sensorsio}). For each tile 60 random spatial patches of size \(64\times 64\) pixels have been randomly generated, from which all available acquisitions between 2016-01-02 and 2024-10-02 have been extracted.

The number of extracted SITS is approximately 4.7 millions, they have a median size of 280 temporal acquisitions, with a maximum and minimum size of 643 and 39 respectively. With time-varying SITS, the objective is to assess the ability of the regularization neural network in homoscedastic and heteroscedastic Whittaker smoothers in such situation.

\begin{figure}
  \centering
  \includegraphics[width=0.75\linewidth]{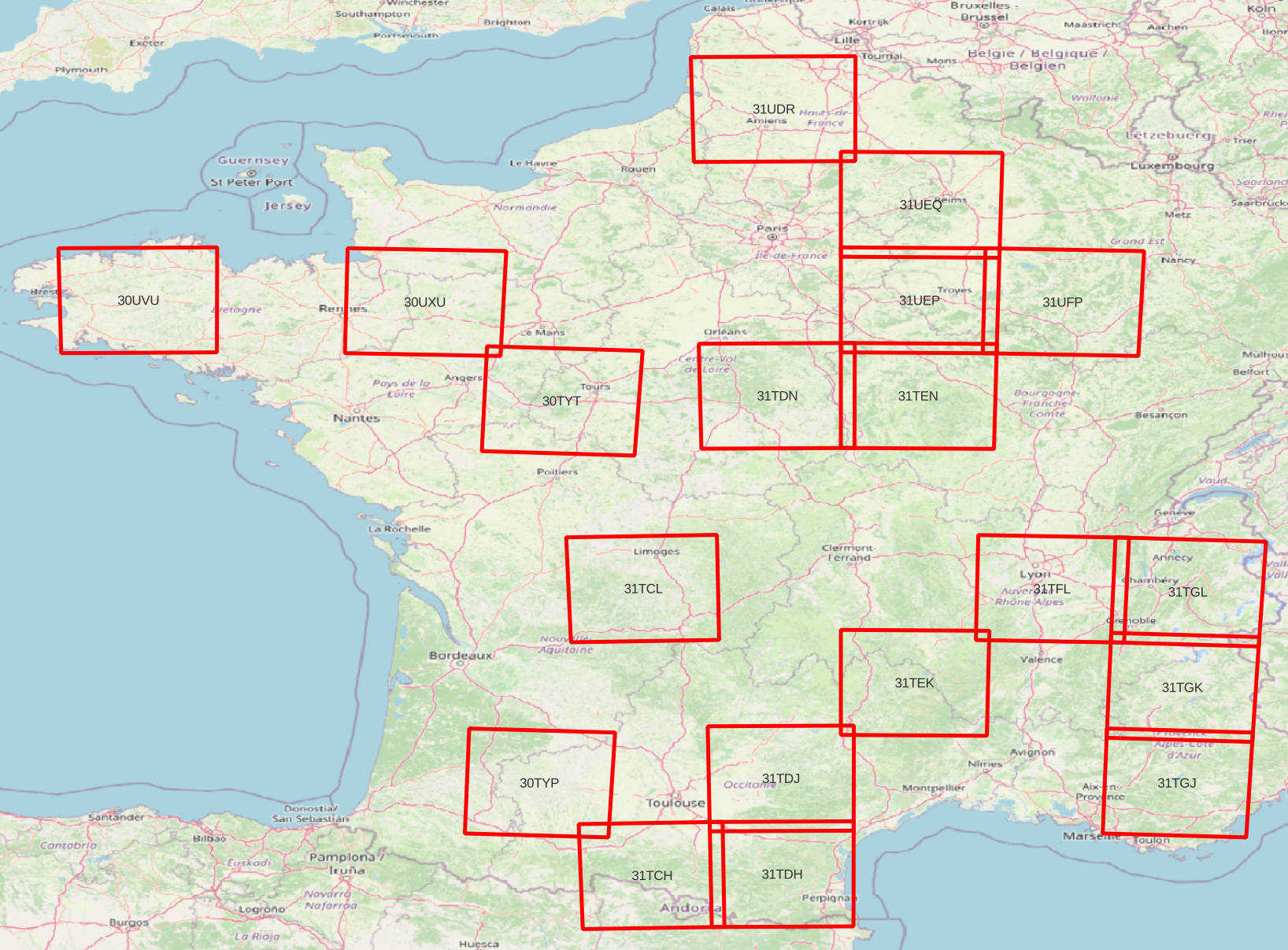}
  \caption{Location of the 19 tiles over the French territory. Background map \textcopyright OpenStreetMap contributors. List of tiles: 31TDJ, 31UDR, 31TCH, 31TGL, 31TGK, 31TCL, 31TDN, 31TFL, 30TYP, 31TGJ, 31TDH, 31UFP, 31UEQ, 31TEN, 31TEK, 31UEP, 30UVU, 30TYT, 30UXU.}
  \label{fig:emprise}
\end{figure}

\subsection{Smoothing parameter}
\label{sec:smoothing:parameter}
As discuss in Section~(\ref{sec:layer}), the heteroscedastic smoothing formulation needs to train a function parameterized by a neural network \(g_{\boldsymbol{\theta}}\). Such neural network should be able to deal with multivariate irregular sequences. To date, the \emph{Transformer} architecture has shown to perform very well with SITS~\cite{MICHEL2026115159} and it was used in this work. However, any architecture satisfying this constraint could have been used.

In these experiments a transformer with sinusoidal position encoding for the time information was used: it takes as input raw SITS and output the smoothing diagonal matrix. To ensure strict positivity, a last sigmoid layer was added ensuring the output of \(g\) ranges between a predefined minimal and maximal value (typically \num{1e-6} and \num{1e10}).

For the homoscedastic case, a scalar value is learned using the same parametrization to ensure positivity and min/max values.

\subsection{Experimental setup}
\label{sec:exp:setup}
A masking strategy was used to optimize \(\boldsymbol{\theta}\) using the original SITS. The mean square error was used as the loss function. Times were masked randomly for each input SITS of the batch: in practice, this means setting the value of \(\mathbf{W}_{tt}\) to zero. For cloud- or edge-induced gaps, the nearest available value (by day) served as the reference. Spectral values were standardized to zero mean and unit variance. The  homoscedastic and heteroscedastic models have been trained for 10 epochs, with a learning rate of 0.0001 and a batch size of 4096 SITS.

It should be noted that for the training or the inference, the cloud mask is not provided to the Whittaker smoother. The objective is to let the regularization network learning the amount of regularization needed for an accurate reconstruction.

\subsection{Results}
\label{sec:results}

\begin{table}
  \centering
  \caption{Reconstruction accuracy. Reported results are averaged over 5 runs. Best results are boldfaced.}
  \label{tab:results}
  \scriptsize
  \begin{tabular}{ccccccc}
    \toprule
    & \multicolumn{3}{c}{MSE} &  \multicolumn{3}{c}{MaxE}\\
    \cline{2-7}
    \diagbox{Model}{Order} & 2 & 3 & 4 & 2 & 3 & 4 \\
    \midrule
    Homoscedastic & 0.30  & 0.37 & 0.32 & 3.82 & 5.14  & 3.09 \\
    \midrule
    Heteroscedastic & 0.31 & 0.39 &\textbf{0.29} & 3.13  & \textbf{3.05}  & 3.86 \\
    \bottomrule
  \end{tabular}
\end{table}
Reconstruction accuracy were estimated using the mean square error (MSE) and the maximum error (MaxE) computed as \(\text{MSE}(\bz, \bx)=T^{-1}\sum_{t=1}^T\bW_{tt}(\bz_t-\bx_t)^2\) and \(\text{MaxE}(\bz,\bx) = \max_{t}\{\bW_{11}|\bz_1-\bx_1|, \ldots,\bW_{TT}|\bz_T-\bx_T|\}\), respectively.

Results are given in Table~\ref{tab:results}. From the table, both approaches provide similar accuracy in terms of MSE and MaxE. Increasing the order of the smoother decreases the accuracy for the homoscedastic model, both in terms of MSE and MaxE. Worst results were obtained for order equal to 3. For the heteroscedastic model, best results were obtained for order equal to 4 for the MSE, but with an increased of the MaxE. 

Training the homoscedastic model took on average 1 hour and 30 minutes while the heteroscedastic model took 3 hours and 20 minutes. The additional complexity comes from the transformer module as well as the higher number of derivatives required. Also, the heteroscedastic model is more demanding in terms of memory consumption for the same reasons: for the same batch, the heteroscedastic model consumes around 25Gb of GPU RAM while the homoscedastic model consumes only around 1Gb. In experiments not included here for brevity, the homoscedastic model was optimized in just 20 minutes by increasing the batch size to 49152 SITS.

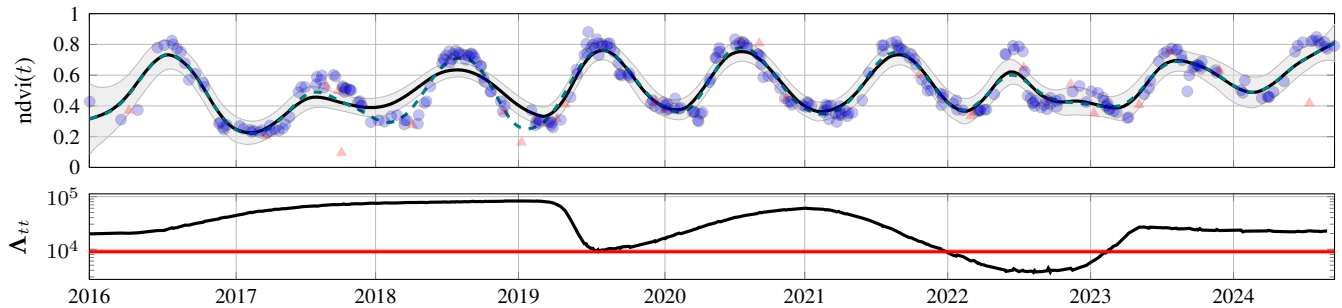
\begin{figure*}
  \centering
  \begin{tikzpicture}
    \pgfplotsset{/pgf/number format/assume math mode=true}
    \begin{groupplot}[group style={group size = 1 by 2, vertical sep=10pt,   x descriptions at=edge bottom, },
      date coordinates in=x, date ZERO=2016-01-01,
      footnotesize,
      xticklabel=\year,
      xticklabel style={
        anchor=near xticklabel,
      },point meta=explicit symbolic,
      xtick={2016-01-01, 2017-01-10,2018-01-01,2019-01-01,2020-01-10,2021-01-01,2022-01-01,2023-01-01,2024-01-01},
      xmin=2016-01-01, xmax=2024-09-15, grid, width=\linewidth, height=0.15\linewidth]
      \nextgroupplot[ymin=0, ymax=1,ylabel={ndvi(\(t\))},height=0.20\linewidth]
      \addplot[scatter, only marks, opacity=0.25, scatter/classes={a={mark=*, fill=blue}, b={mark=triangle*, red}}] table [x=doy, y=ndvi, meta=mask, col sep=comma] {figures/ndvi_scalar.csv};

      \addplot[name path=stdp, mark=None,gray!50] table [x=doy, y expr={\thisrow{ndvi_pred}+4*\thisrow{ndvi_pred_var}}, col sep=comma] {figures/ndvi_pred_batch_multi.csv};
      \addplot[name path=stdm, mark=None,gray!50] table [x=doy, y expr={\thisrow{ndvi_pred}-4*\thisrow{ndvi_pred_var}}, col sep=comma] {figures/ndvi_pred_batch_multi.csv};
      \addplot[gray!25, opacity=0.5] fill between[of=stdp and stdm];
      \addplot[black, very thick] table [x=doy, y=ndvi_pred, col sep=comma] {figures/ndvi_pred_batch_multi.csv};
      
      \addplot[teal, very thick, dashed] table [x=doy, y=ndvi_pred, col sep=comma] {figures/ndvi_pred_scalar.csv};

      \nextgroupplot[ymode=log, ylabel={\(\bL_{tt}\)},height=0.15\linewidth]
      \addplot[black, very thick] table[x=doy, y=reg_1, col sep = comma] {figures/reg_val.csv};
      \addplot[red, very thick] coordinates{(2016-01-01, 9000) (2024-09-15, 9000)};
    \end{groupplot}
  \end{tikzpicture}
  \caption{Reconstruction of NDVI SITS. In the upper plot, colored dots and triangles represent the raw acquisitions between 2016 and 2024, dot are valid dates and triangle are non valid dates. The black continuous curve represents the reconstruction obtained with the heteroscedastic model, while the shaded area indicates the credibility intervals calculated using equation~(2.2) from~\cite{biessy:hal-04124043}. The dashed teal curve is the reconstruction using the homoscedastic model. In the lower plot, red and black curves represent the smoothing parameters for homoscedastic and heteroscedastic models, respectively.}
  \label{fig:plot:sits}
\end{figure*}

The figure~\ref{fig:plot:sits} shows the reconstruction of a SITS with a stepsize of 5 days for the whole temporal period for the two models as well as their smoothing parameter estimates. To ease analyse, the normalized difference vegetation index (NDVI) was computed from the raw and the smooth SITS. From the plot below, it can be seen that the heteroscedastic model learn a varying smoothing parameter along the time. Yet, the resulting smoothing SITS matchs almost all the time the one obtained with homoscedastic model. Differences are more noticeable in regions with higher smoothing parameter values and lower temporal data acquisition frequency, such as between 2018 or 2019. Elsewhere, both model produces very similar reconstruction.

\section{Discussion and conclusion}
\label{conclusion}
This paper introduced the Whittaker smoother as a neural network layer, enabling the optimization of the smoothing parameter via stochastic gradient descent. A sparse, efficient, and differentiable implementation of the associated inverse problem was developed to handle large-scale datasets. Experimental results demonstrated that this implementation substantially outperforms standard linear solvers in terms of processing time and memory consumption. Large-scale satellite image time series (SITS) were processed using both homoscedastic and heteroscedastic formulations of the proposed method — marking the first application of a heteroscedastic Whittaker filter at such scale. While results confirmed the feasibility of the heteroscedastic approach, the performance gap in terms of reconstruction error with the homoscedastic model remained limited. This may be attributed to the transformer network used to estimate the smoothing parameter in the heteroscedastic setting, whose predictions appear insufficiently sensitive to rapid noise variations along a SITS — such as cloud cover that can appear and vanish within a single day. A promising direction for future work would be to explore neural architectures better suited to capturing sharp temporal features, such as the SIREN  architecture~\cite{DBLP:journals/corr/abs-2006-09661}.

\section*{Acknowledgement}
The author wishes to express its gratitude to J.~Inglada and J.~Michel (CESBIO lab) for their valuable insights regarding the implementation and experimental design, and to B.~Charlier (MIA Toulouse) for enriching discussions on the conjugate gradient method.
\AtNextBibliography{\small}
\printbibliography
\end{document}